\documentclass[aps, pra,twocolumn,reprint, longbibliography, superscriptaddress,showpacs]{revtex4-1}

\usepackage{amsmath}
\usepackage{amssymb}
\usepackage{enumitem}
\usepackage{young}
\usepackage{ytableau}
\usepackage{graphicx}
\usepackage{youngtab}
\usepackage[hidelinks=true]{hyperref}
\usepackage{color}
\usepackage{kantlipsum,setspace}
\makeindex

\newcommand{\su}[2]{\mathrm{su}\!\left(#1\right)_{#2}}
\newcommand{\nsu}[2]{\mathrm{su}\left(#1\right)_{#2}}
\newcommand{\pf}[3]{\mathrm{su}\!\left(#1\right)_{#2}/\mathrm{u}\!\left(1\right)^{#3}}
\newcommand{\bg}{\mathcal{O}_{\mathrm{bg}}}

\begin{document}
\begin{abstract}
The fractional quantum Hall effect is the paradigmatic example of topologically ordered phases.
One of its most fascinating aspects is the large variety of different topological orders that may be realized, in particular nonabelian ones. 
Here we analyze a class of nonabelian fractional quantum Hall model states which are generalizations of the abelian Haldane-Halperin hierarchy.
We derive their topological properties and show that the
quasiparticles obey nonabelian fusion rules of type $\su{q}{k}$.
For a subset of these states we are able to derive the conformal field theory description that makes the topological properties --- in particular braiding --- of the state manifest. 
The model states we study provide explicit wave functions for a large variety of interesting topological orders, which may be relevant for certain fractional quantum Hall states observed in the first excited Landau level.  
\end{abstract}

\title{Conformal field theory construction for nonabelian hierarchy wave functions}

\newcommand{\col}{Institut f{\"u}r Theoretische Physik,
Universit{\"a}t zu K{\"o}ln, D-50937 Cologne, Germany}

\newcommand{\gu}{Department of Physics, University of Gothenburg, SE 412 96 Gothenburg, Sweden}

\author{Yoran Tournois}
\author{Maria Hermanns}
\affiliation{\col}
\affiliation{\gu}

\maketitle
\section{Introduction}
\label{sec:intro}

Topological phases of matter are among the most fascinating phases in condensed matter systems.
A unifying feature of such phases is that certain properties are robust and independent of microscopic details.
Arguably the most fascinating example of a topological phase are the fractional quantum Hall liquids \cite{Tsui1982two} --- both because of the vast variety of quantum liquids that have been observed and their exotic quasiparticles, which obey anyonic statistics interpolating between bosonic and fermionic statistics \cite{Leinaas1977theory}. 

Most of the progress within the fractional quantum Hall (FQH) effect has been made by proposing simple trial wave functions that capture the topological properties of the topological phase. 
This strategy was pioneered by Laughlin \cite{laughlintheory},
who conjectured that the FQH liquids at filling fractions $\nu=1/m$ have quasiparticle excitations with fractional electric charge $e/m$. 
Laughlin's idea was generalized by Haldane \cite{Haldane1983fractional} 
and Halperin \cite{Halperin1984statistics}, who conjectured that all FQH liquids in the lowest Landau level could be understood using a simple hierarchical picture, where new FQH liquids may be formed by successive condensation of quasiparticles (see Ref. \cite{Hansson2017quantum} for a comprehensive review on this topic). 
This picture gives a systematic understanding of the physics in the lowest Landau level (LLL) --- it not only predicts that FQH liquids may occur at any filling fraction $\nu=p/q$ with $p,q$ relatively prime and $q$ odd, but also estimates their relative stability $\sim 1/q$ and determines the fractional charge of the quasiparticles to be $\pm e/q$. 

In the hierarchical picture trial wave functions at hierarchical level $n+1$ can be obtained from those at level $n$ by a recursion formula  \cite{Halperin}. However, this recursion formula can only be evaluated explicitly for a subset of states, namely those obtained by successive condensation of quasi\emph{electrons} (as opposed to quasiholes) \cite{Hansson2009quantum}. 
In this case the wave functions turn out to be identical to the composite fermion (CF) wave functions \cite{Jain1989composite}.
The latter have been extensively tested numerically -- for small system sizes -- against the exact ground state for Coulomb interaction, verifying that these states indeed capture the topological properties of the low-energy sector (see for instance \cite{Jainbook} and references therein).

Conformal field theory (CFT) has proven to be a useful tool for proposing and analyzing model wave functions. It was pointed out by Moore and Read \cite{MooreRead} and Fubini \cite{Fubini1991vertex} that the Laughlin wave function can be written as a correlator of vertex operators in the chiral boson CFT. Moore and Read conjectured that other CFTs could also be used to propose valid FQH model states. In this picture, electrons and quasiholes are represented by local primary operators in the CFT. 
These encode the topological properties of the particles such as fractional electric charge as well as fusion and braiding properties. 
This approach was generalized in various ways.  
Firstly, one can identify a quasi-local CFT operator that creates quasielectron excitations for any model state that can be written as a correlator of a CFT \cite{QuasielectronCFT,Hansson2009quantum}.
Secondly, using several copies of chiral boson fields one can describe the (spinful) Halperin $\left(mmn\right)$ states \cite{Halperin,MooreRead} and the Haldane-Halperin hierarchy/composite fermion states \cite{CFTCF,Hansson2007conformal,Hansson2009quantum}. 
Finally, one can also combine the chiral boson with slightly more complicated CFTs --- e.g. parafermion theories \cite{Zamolodchikov1985nonlocal} for the Read-Rezayi series \cite{ReadRezayi} or Gepner parafermions \cite{GepnerParafermions} for the nonabelian spin-singlet (NASS) state \cite{NASS} and its generalizations \cite{NASSRR,Reijnders2002quantum,FujiLecheminant}. 
These latter model states may be relevant for describing quantum Hall liquids occurring in the first excited Landau level. 

The physics of the lowest and first excited Landau level is, in fact, quite different. 
This became evident with the experimental discovery of an incompressible state at filling fraction $5/2$, which was the first quantum Hall liquid with an even denominator. 
This state is believed (see e.g. \cite{Rezayi2017landau} and references therein) to be in the same topological phase as the `anti-Pfaffian' \cite{AntiPfaffian1,AntiPfaffian2} --- the particle-hole conjugate of the Moore-Read (or Pfaffian) state \cite{MooreRead} --- and is  expected to host nonabelian quasiparticle excitations that behave as Ising anyons \cite{Belavin1984infinite}.
The $5/2$ state is the first, but not the only state that is believed to harbor nonabelian anyons in the first excited Landau level. 
In fact, numerical simulations \cite{wojs2001electron} suggest that  pairing of electrons is favorable in the first excited Landau level, which in turn can stabilize nonabelian FQH liquids. 
There are several experimentally observed filling fractions that are believed to fall into this class, but the  most prominent example --- apart from the Moore-Read state --- is arguably the FQH liquid at filling fraction $12/5$ \cite{panLL2}. 
Currently, it is most commonly believed to be in the same topological phase as the particle-hole conjugated $\mathbb{Z}_3$ Read Rezayi state, harboring Fibonacci anyons \cite{ReadRezayi}. 
But numerical simulations have proven to be hard \cite{Mong2017fibonacci,Pakrouski2016enigmatic} and there are several possible candidate states with very different predictions on the nonabelian nature of the state, such as Ising anyons \cite{Bonderson2008fractional} and $\su{4}{2}$-type anyons \cite{NAHierarchy}. 

One may wonder if there is a  systematic way to understand the observed filling fractions in the first excited Landau level, similar to the Haldane-Halperin hierarchy in the lowest Landau level. There have been several attempts to generalize this hierarchy to nonabelian liquids, most notably by Bonderson and Slingerland \cite{Bonderson2008fractional},  Levin and Halperin \cite{Levin2009collective}, and one of the authors \cite{NAHierarchy}. 
In this manuscript we focus on the last, because it is the only one of the three proposals that allows for a nonabelian CFT description that is richer than the Ising CFT.
In particular, it suggests a series of \emph{spin-polarized states} with $su(q)_k$ type statistics; other realizations of such type of statistics rely on spinful (i.e. spin-unpolarized) particles \cite{NASS, NASSRR,Reijnders2002quantum,FujiLecheminant} or assume some other type of `flavor' quantum number \cite{Wen1991statistics,BlokWen1992}.
This hierarchy can be constructed by successive condensation of nonabelian quasiparticles,  in a very similar way to that proposed by Halperin for abelian states \cite{Halperin}. 
The resulting wave functions can be brought into a particularly simple form, which for bosons reduces to
\begin{align}
\label{eq:NAhier}
\Psi_{\nu}^{k}=\mathcal{S}\big[\underbrace{\Psi_{\nu}\ldots\Psi_{\nu}}_{k\text{ copies}}\big]
\end{align}
where one symmetrizes over $k$ copies of the abelian, bosonic wave functions $\Psi_{\nu}$ of Haldane-Halperin or composite fermion type at filling fraction $\nu=p/q$. 

This paper is organized as follows. In Section \ref{sec:na} we introduce generalized Halperin and clustered spin-singlet states, after which we introduce the nonabelian hierarchy wave functions. For a certain subset of these states, we determine the explicit CFT description, in which all the topological properties should be manifest. In Section~\ref{sec:tt} we study the full set of nonabelian hierarchy wave functions in the thin torus limit. In this simplifying limit, we determine the fusion rules of the quasiparticle excitations. 
To illustrate our methods, we present an explicit example in Section~\ref{sec:ex}.
We conclude our discussions in Section~\ref{sec:conclusion}. In Appendix~\ref{sec:roots}, we provide details on the choice of roots and weights in writing the clustered spin-singlet and nonabelian hierarchy states. In Appendix~\ref{sec:para}, we provide a brief introduction to parafermion CFTs and give some details on the parafermion theory $\pf{4}{2}{3}$ used in Section~\ref{sec:ex}. Finally, we determine the quasiparticle degeneracy for a subset of nonabelian hierarchy wave functions in Appendix~\ref{sec:qpdeg}.

\paragraph*{Notation:} We denote the particle coordinates by $z=x+iy$ and quasiparticle coordinates by $\eta$. They may carry an subscript $\sigma$ which refers to (pseudo)spin, or a superscript $a$ which is a layer index. $\{z_{i}\}$ and $\{z_{i,\sigma}\}$ always refer to all particle coordinates and similarly for $\{\eta_{i}\}$ and  $\{\eta_{i,\sigma}\}$, whereas $\{z_{i,\sigma}^a\}$ refers to all particles of layer $a$.
We label the nonabelian hierarchy states by a superscript $k$ that denotes the number of copies or `layers' and a subscript $\nu$ that denotes the filling fraction in each of the layers. Note that the filling fraction of such a nonabelian state is given by $k \nu$. When $k=1$, we often suppress the superscript and label the state only by its filling fraction. 
The various spin-singlet states are labeled by the algebra they are based on. 
We suppress Gaussian factors throughout. 

\section{nonabelian hierarchy wave functions}
\label{sec:na}
In this section we introduce the nonabelian hierarchy wave functions. In order to set up notation and introduce important concepts we first review several other model states and discuss their interconnections. In the remainder of the manuscript we only consider bosonic wave functions, as it simplifies the discussion. Note that the fermionic wave functions can always be obtained by multiplying by a full Jastrow factor. These are expected to have the same nonabelian properties --- only abelian phases (connected to the charge of the particles) change. 
By a slight abuse of notation, we still call the bosonic particles `electrons' as they correspond to the electron operators for the corresponding fermionic states, which we are ultimately interested in. Further, we will refer to quasiholes and quasielectrons collectively as quasiparticles. 
\paragraph*{Haldane-Halperin hierarchy and composite fermion states}
The Haldane-Halperin hierarchy is a simple picture to explain the zoo of filling fractions in the lowest Landau level and their properties \cite{Haldane1983fractional,Halperin1984statistics}. 
Starting from the Laughlin states and successively condensing quasiparticle excitations one obtains FQH liquids  at any filling fraction $p/q$, where $q$ is odd and $p$ and $q$ are relatively prime. 
Their quasiparticle excitations have fractional charge $\pm e/q$, which determines the relative stability of FQH liquids to be roughly given by $1/q$. 
Halperin also proposed a recursion formula for explicit wave functions at hierarchy level $n+1$ given a parent state at level $n$. It reads
\begin{align}
\label{eq:HH}
\Psi_{\nu_{n+1}}(\{z_i\}) =&   \prod_{j=1}^{N_p} \int  d^2\vec \eta_j  \, \Phi^\star(\{\eta_j\})  \Psi_{\nu_n} (\{\eta_j\} ;\{z_j\})  \, ,
\end{align}
where $\Psi_{\nu_n} (\{\eta_j\} ;\{z_j\})$ is a parent state with $N_p$ quasiparticles at positions $\eta_1,\ldots, \eta_{N_p}$ and $\Phi$ is a pseudo wave function (a wave function of the quasiparticles), which for abelian states is a Laughlin wave function. 
The wave functions \eqref{eq:HH} are generically hard to evaluate, even for small numbers of particles. 
However, for the subset of states that arise from condensing quasielectrons only, the integrals can be performed analytically \cite{Hansson2009quantum}. 
The resulting wave functions are identical to those of composite fermions \cite{Jain1989composite}, whenever the latter exist. 
\paragraph*{Generalized Halperin states}
The bosonic Halperin $\left(221\right)$ spin-singlet wave function \cite{Halperin} is given by
\begin{equation}
\begin{aligned}
\Psi_{\nsu{3}{1}}\! \left(\left\{z_{i,\sigma}\right\}\right) &= \prod_{i<j}^{N_{\uparrow}} \left(z_{i,\uparrow}-z_{j,\uparrow}\right)^2 \prod_{i<j}^{N_{\downarrow}} \left(z_{i,\downarrow}-z_{j,\downarrow}\right)^2 \\
&\times\prod_{i=1}^{N_\uparrow} \prod_{j=1}^{N_\downarrow} \left(z_{i,\uparrow}-z_{j,\downarrow}\right)^1,
\label{eq:HalpWF}
\end{aligned}
\end{equation}
where $\sigma=\uparrow,\downarrow$ is the (pseudo)spin index. The wave function is denoted by $\Psi_{\nsu{3}{1}}$ as it has an $\su{3}{1}$ symmetry \cite{NASS} and it is an $\su{2}{}$ singlet when the number of particles per group is equal, $N_{\uparrow}=N_{\downarrow}$. By allowing the pseudospin index $\sigma$ to have values $1,\ldots,n$ we obtain the generalized Halperin state
\begin{multline}
\Psi_{\nsu{n+1}{1}}\!\left(\left\{z_{i,\sigma}\right\}\right) = \prod_{\sigma=1}^{n} \prod_{i<j}^{N_{\sigma}} \left(z_{i,\sigma} -z_{j,\sigma}\right)^2 \\ \times \prod_{\sigma<\sigma'}^{n} \prod_{i=1}^{N_\sigma}\prod_{j=1}^{N_{\sigma'}} \left(z_{i,\sigma}- z_{j,\sigma'}\right)^{1},
\label{eq:ASUN}
\end{multline}
which is an $\su{2}{}$ singlet when all $n$ groups have equal size $N_\sigma = M$. 
The filling fraction is $\nu=\frac{n}{n+1}$, and the wave function has an $\su{n+1}{1}$ symmetry \cite{FujiLecheminant}. 
Quasihole wave functions for these states are obtained by inserting Laughlin quasiholes in any of the pseudospin groups; if there are $Q_\sigma$ quasiholes in pseudospin group $\sigma$, the wave function reads
\begin{multline}
\label{eq:QHWF}
\Psi_{\nsu{n+1}{1}}\!\left(\left\{\eta_{i,\sigma}\right\};\left\{z_{j,\sigma'} \right\} \right) = \\ \prod_{\sigma=1}^{n} \prod_{i=1}^{Q_\sigma}\prod_{j=1}^{N_\sigma} \left(\eta_{i,\sigma} - z_{j,\sigma} \right) \Psi_{\nsu{n+1}{1}}\! \left(\left\{z_{j,\sigma'}\right\}\right).
\end{multline}
The $\su{n+1}{1}$ symmetry of the generalized Halperin states may be used to express them as conformal blocks; they can be written
\begin{equation}
\Psi_{\nsu{n+1}{1}}\! \left(\left\{z_{i,\sigma}\right\}\right)= \left<\prod_{i=1}^{M} V_{\tilde{\alpha}_1}\! \left(z_{i,1}\right)  \cdots \prod_{i=1}^{M} V_{\tilde{\alpha}_n} \!\left(z_{i,n}\right)\bg\right>,
\label{eq:ASUNCorr}
\end{equation}
in terms of $n$ types of electron operators $V$ --- one for each value of pseudospin. The electron operators are vertex operators of $n$ independent chiral bosonic fields 
\begin{equation}
V_{\tilde{\alpha}}\!\left(z\right)={:}e^{ i \tilde{\alpha} \cdot \vec{\varphi} \left(z\right)}{:}
\label{eq:AVertOp}
\end{equation}
with $\vec{\varphi}\!\!\!=\left(\varphi_1,...,\varphi_n\right)$ and $\left<\varphi_i \left(z\right) \varphi_j \left(w\right)\right> = - \delta_{ij} \ln \left(z-w\right)$. The operator $\bg$ is a background charge operator, which is inserted in the correlator to ensure charge neutrality \cite{MooreRead}. Finally, the vertex operator is labeled by a vector $\tilde{\alpha}$, which may be identified as a root of $\su{n+1}{}$. The correlator in Eq. \eqref{eq:ASUNCorr} reproduces the wave function Eq. \eqref{eq:ASUN} if we choose $\tilde{\alpha}_{i}\cdot \tilde{\alpha}_i =2$ for all $i$ and $\tilde{\alpha}_i\cdot \tilde{\alpha}_j=1$ for $i\neq j$. We will make the choice
\begin{equation}
\tilde{\alpha}_i = \sum_{j=1}^{i} \alpha_j,
\label{eq:rootseq}
\end{equation}
where $\alpha_j$ are the simple roots of $\su{n+1}{}$. This still leaves freedom for the explicit choice of the roots $\tilde{\alpha}_1,\ldots,\tilde{\alpha}_n$,  which can be exploited to ensure that $\bg$ only depends on one field, see Appendix \ref{sec:roots}. 
The quasihole wave functions \eqref{eq:QHWF} may also be written as a CFT correlator by inserting vertex operators that represent the quasiholes. There are $n$ types of quasiholes and the relevant vertex operators are
\begin{equation}
H_{\tilde{\omega}_i} \!\left(\eta\right)  = {:}e^{i \tilde{\omega}_i \cdot \vec{\varphi}\left( \eta\right)}{:}
\label{eq:AQh}
\end{equation}
with $i=1,...,n$. Here $\tilde{\omega}_i=\omega_{i}-\omega_{i+1}$ for $i\leq n-1$ and $\tilde{\omega}_n=\omega_n$, with $\omega_i$ the fundamental weights of $\su{n+1}{}$. 
The $\tilde{\omega}_i$ are dual to the roots in Eq. \eqref{eq:rootseq}, i.e. $\tilde{\alpha}_i \cdot \tilde{\omega}_j=\delta_{ij}$, which ensures that the operators \eqref{eq:AQh} amount to inserting a quasihole in the $i$-th pseudospin group.

The generalized Halperin states $\Psi_{\nsu{n+1}{1}}$ are closely related to the (bosonic) positive Jain series. 
Introducing derivatives for each pseudospin group in Eq.~\eqref{eq:ASUN}
\begin{align}
\partial_{\sigma}&\equiv \prod_{j=1}^{M}\frac{\partial}{\partial z_{j,\sigma}}
\label{eq:der}
\end{align}
and symmetrizing over the pseudospins afterwards,  one reproduces the positive Jain series identically \cite{Hansson2009quantum}
\begin{align}
\label{eq:SUnCF}
 \Psi_{n/n+1}(\{z_j\})&=\mathcal{S}\left[\left(\prod_{\sigma=1}^{n} \partial_{\sigma}^{\sigma-1}\right)\Psi_{\nsu{n+1}{1}}\!\left(\left\{z_{i,\sigma}\right\}\right)\right].
\end{align}
This does not only hold for the ground states, but for all the quasiparticle excitations as well. 
It should be noted that the derivatives do not act on the (suppressed) Gaussian part of the wave function in Eq. \eqref{eq:SUnCF}. 

\paragraph*{Clustered spin-singlet states}
Originally, nonabelian spin-singlet (NASS) states \cite{NASS,NASSRR} were proposed as generalizations of the (spin-polarized) Read-Rezayi (RR) series \cite{ReadRezayi}. 
The RR states are based on $\su{2}{k}$ and can be written as CFT correlators involving a product of neutral $Z_k$ parafermions with a vertex operator. 
The NASS states are spin-singlets and their formulation is in terms of conformal blocks based on the algebra $\su{3}{k}$ for $k>1$,
\begin{equation}
\Psi_{\nsu{3}{k}}\!\left(\{z_{i,\sigma}\}\right)= \left<\prod_{i=1}^{M} V_{\tilde{\alpha}_1} \left(z_{i,\uparrow}\right) \prod_{i=1}^{M} V_{\tilde{\alpha}_2} \left(z_{i,\downarrow}\right)\bg\right>.
\label{eq:NASSCorr}
\end{equation}
Here, the electron operators
\begin{equation}
V_{\tilde{\alpha}} \!\left(z\right) = \psi_{\tilde{\alpha}} \! \left(z\right){:}e^{i\tilde{\alpha} \cdot \vec{\varphi} \left(z\right)/\sqrt{k}}{:}
\label{eq:NASSV}
\end{equation}  
 factorize into a neutral Gepner parafermion $\psi_{\tilde{\alpha}}$ of $\pf{3}{k}{2}$ (see Appendix \ref{sec:para} for a short introduction to Gepner parafermions) and a vertex operator in terms of roots $\tilde{\alpha}_1,\tilde{\alpha}_2$ of $\su{3}{}$. One obtains quasihole wave functions by inserting operators which likewise split into a neutral and a charged part. There are two quasihole operators, one for each pseudospin component,  given by 
\begin{equation}
H_{\tilde{\omega}_i} \!\left(\eta\right) = \sigma_{\tilde{\omega}_i} \!\left(\eta\right) e^{i \tilde{\omega}_i \cdot \vec{\varphi} \left(\eta\right)/\sqrt{k}},
\label{eq:NASSH}
\end{equation}
where $\sigma_{\tilde{\omega}_i}$ are primary fields in the parafermion theory.

By allowing the (pseudo)spin index to have $n$ values, one obtains a clustered spin-singlet state \cite{Reijnders2002quantum,FujiLecheminant}, which may be expressed as 
\begin{equation}
\Psi_{\nsu{n+1}{k}}\! \left(\{z_{i,\sigma}\}\right) = \left< \prod_{i=1}^{M} V_{\tilde{\alpha}_1}\!\left(z_{i,1}\right) \cdots \prod_{i=1}^{M} V_{\tilde{\alpha}_n}\! \left(z_{i,n}\right)\bg\right>.
\label{eq:NASUNCorr}
\end{equation}
The electron operators again have the form as in Eq. \eqref{eq:NASSV}, with $\psi_{\tilde{\alpha}}$ a Gepner parafermion of $\pf{n+1}{k}{n}$. The roots $\tilde{\alpha}_1,\ldots,\tilde{\alpha}_n$ can be chosen as in Eq. \eqref{eq:rootseq} \footnote{Note that this choice is different from the one made in \cite{NASS}, although it is equivalent}. Finally, the quasihole operators generalize in a similar way: they are given as in Eq. \eqref{eq:NASSH} where the fields $\sigma_{\tilde{\omega}_i}$ are primary fields in the parafermion theory $\pf{n+1}{k}{n}$ and $\tilde{\omega}_i$  is given as in the abelian case. 

Alternatively, the wave function $\Psi_{\nsu{n+1}{k}}$ may be expressed as a symmetrizer over $k$ copies of the abelian generalized Halperin state, similar to the representation of the Read-Rezayi states proposed by Cappelli, Georgiev, and Todorov \cite{Cappelli}. 
They showed that both the ground state and excited states of the RR series can be rewritten as symmetrizers over  bosonic $\nu=\frac{1}{2}$ Laughlin states \cite{laughlintheory}. The ground state is given by 
\begin{equation}
\Psi_{\nsu{2}{k}} \!\left(\{z_{i}\}\right)= \mathcal{S} \left[ \prod_{a=1}^k \prod_{i<j}\left(z_{i}^{a}-z_{j}^{a}\right)^2  \right],
\label{eq:RR}
\end{equation} 
where the $N$ particle coordinates are divided into $k$ `layers' of size $N/k$. The full wave function is obtained by symmetrizing over all ways of dividing the coordinates between the different layers. 
Similarly the NASS state in Eq. \eqref{eq:NASSCorr} can be expressed as a symmetrizer over $k$ copies of the Halperin $\left(221\right)$ wave functions and for general $n$, we have 
\begin{equation}
\Psi_{\nsu{n+1}{k}}\!\left(\left\{z_{i,\sigma}\right\}\right)= \mathcal{S}_{l} \left[ \prod_{a=1}^{k} \Psi_{\nsu{n+1}{1}}^{a} \!\left(\left\{z_{i,\sigma}^{a} \right\}\right)\right].
\label{eq:NASUNSym}
\end{equation}
Here each coordinate is labeled by a pseudospin index $\sigma$, a layer index $a$, as well an index $i$. $\{z_{i,\sigma}^a\}$ denotes the set of all particles in layer $a$ and the symmetrization $\mathcal{S}_l$ is a partial symmetrization which symmetrizes over all layers, but not the pseudospins. That is, it symmetrizes over all inequivalent ways of dividing the coordinates with a given pseudospin over the layers.

\paragraph*{Nonabelian hierarchy wave functions}
\label{sec:NAH}
Now we have all the ingredients necessary to realize that the nonabelian hierarchy wave functions in  Eq.~\eqref{eq:NAhier} are rather natural generalizations of the clustered spin-singlet states. 
Instead of symmetrizing over $k$ layers of abelian generalized Halperin states as done in Eq.~\eqref{eq:NASUNSym}, we symmetrize over $k$ layers of composite fermion states Eq. \eqref{eq:SUnCF}. By performing the partial symmetrization over the layers first one obtains
\begin{equation}
\label{eq:SU(n)_kSym}
\Psi_{n/n+1}^{k}= \mathcal{S}\left[ \partial_1^0 \partial_{2}^{1} \cdots \partial_{n}^{n-1} \Psi_{\nsu{n+1}{k}}\!\left( \left\{z_{i,\sigma}\right\}\right) \right],
\end{equation}
where the symmetrization $\mathcal{S}$ is over the pseudospin. Using the explicit CFT description of the clustered spin-singlet states we obtain a (nearly) explicit CFT description of the nonabelian hierarchy wave functions
\begin{equation}
\Psi_{n/n+1}^{k} = \mathcal{S} \left< \prod_{i=1}^{M}\tilde{V}_{\tilde{\alpha}_1}\! \left(z_{i,1}\right)  \cdots \prod_{i=1}^{M} \tilde{V}_{\tilde{\alpha}_{n}} \!\left(z_{i,n}\right) \bg\right>,
\label{eq:NAMaxDensCorr}
\end{equation}
where  $\tilde{V}_{\tilde{\alpha}_j}\! \left(z\right) = \left(\partial_z\right)^{j-1} V_{\tilde{\alpha}_j}\! \left(z\right)$. 
Note that the symmetrization here has a very different effect than that of Eq.~\eqref{eq:NASUNSym}. 
In the latter, we symmetrized over indistinguishable particles, thus reverting a multi-layered abelian state into a nonabelian one. 
Here, we symmetrize over distinguishable particles which differ by their orbital spin. 
This symmetrization is commonly believed not to change the (nonabelian) properties of the state, thus the nonabelian hierarchy state should have the same fusion and braiding properties as the clustered spin-singlet state. 
For the fusion rules we can prove this explicitly using the thin torus analysis. The fusion rules are subsequently used to determine the quasiparticle degeneracy in Appendix \ref{sec:qpdeg}. 

\section{Fusion rules using the thin torus limit}
\label{sec:tt}
\subsection{Thin torus limit}
Certain topological properties of fractional quantum Hall states can be determined in a simple manner by placing the states on a torus and considering the limit where one circumference becomes very small. These properties include the topological ground state degeneracy, the charge of the fundamental quasiparticles, and their fusion rules \cite{TTLimit,Seidel2005incompressible,Bergholtz2006pfaffian,Seidel2006abelian,Ardonne2008degeneracy,ArdonneFusion}. 

In the Landau gauge on the torus the single particle orbitals are Gaussians localized on rings around one of the handles of the torus, which are separated by $\frac{2\pi \ell^2}{L}$ with $L$ the circumference of the torus and $\ell$ the magnetic length. As the thickness $L\to0$ hopping between orbitals is exponentially suppressed and the  degenerate ground states on the torus become simple patterns of occupation numbers. These patterns are determined by the filling fraction and the electrostatic repulsion between the constituent particles. As an explicit example we consider the thin torus limit of the wave function $\Psi_{3/4}$ obtained by letting $n=3$ and $k=1$ in Eq. \eqref{eq:SUnCF}. The filling fraction is $\nu=\frac{3}{4}$, which leads to four degenerate ground state patterns \cite{TTLimit}
\begin{equation}\label{eq:abfusion}
\begin{aligned}
\cdots  01110111\cdots &= \left[0111\right] \\
 \cdots 10111011\cdots &= \left[1011\right]\\
 \cdots 11011101\cdots &= \left[1101\right]\\
 \cdots 11101110\cdots &= \left[1110\right].
\end{aligned}
\end{equation}
Here we have represented the patterns of occupation numbers by their `unit cells'. 
These obey the rule that every four consecutive orbitals contain precisely three particles \cite{TTLimit}.

Excitations correspond to local violations  of this rule, i.e. they are domain walls between degenerate ground state sectors. Here we are mostly interested in the fundamental quasiparticles,  which correspond to domain walls that carry a minimal deficit or excess charge. If $\nu=\frac{p}{q}$, these domain walls are located at strings of $q$ consecutive orbitals with $p\pm1$ charge. In the above example a domain wall structure of the quasielectron is
\begin{equation*}\label{eq:abqe}
\cdots |0111|0\underline{111|1}011|1011| \cdots, 
\end{equation*}
which we denote by $\left[0111\right]\to\left[1011\right]$. Note that the underlined region is the only string of $4$ consecutive orbitals containing $4$ charges, while all other such strings have $3$ charges. Inserting these domain walls at four well-separated positions and comparing to the ground state one can show that this region carries an excess charge $\frac{e}{4}$ \cite{Su81fractionally}. 
The vertical bars are a guide to the eye and show that upon inserting the quasielectron the sector $\left[0111\right]$ is connected to the sector $\left[1011\right]$. Repeating this procedure we obtain the domain wall structures
\begin{equation}\label{eq:abelianfusion}
\begin{aligned}
\left[0111\right] &\to \left[1011\right]\\
\left[1011\right] &\to \left[1101\right]\\
\left[1101\right] &\to \left[1110\right]\\
\left[1110\right] &\to \left[0111\right].
\end{aligned}
\end{equation}
There is a simple `hopping rule' that determines which ground state patterns can be connected by the insertion of a fundamental quasihole/quasielectron. 
When inserting a quasielectron the unit cell of the final ground state sector is found by hopping an electron one site to the left in the unit cell of the original sector, imposing periodic boundary conditions on the unit cell, and disallowing multiple occupation of the orbitals. For a quasihole, the electron has to hop one site to the right instead. 
More details on the thin torus limit can be found in Ref. \onlinecite{TTLimit}. 

\subsection{Fusion rules}
We now use these domain wall structures to determine the fusion rules of the nonabelian hierarchy states, following the method introduced in Refs. ~\cite{Ardonne2008degeneracy,ArdonneFusion}. 
In order to familiarize ourselves with the method, let us start with two simple examples for which the fusion rules are well known. 
 
We first consider the fusion rules for the abelian hierarchy wave functions. 
From the example of $\nu=3/4$ mentioned earlier, it is evident that by successively inserting fundamental quasiparticles, one moves cyclically through the $q=4$ distinct unit cells, as shown in Eq. \eqref{eq:abelianfusion}. 
Using the results of Ref.~\cite{TTLimit}, it is straightforward to show that this generalizes to generic abelian hierarchy states and that the fusion rules depend only on the denominator of the filling fraction. 
In particular an abelian hierarchy state at filling $p/q$ harbors quasiparticles with fusion rules $\su{q}{1}= \mathbb{Z}_q$.

Another well-known example is the bosonic Moore-Read state, obtained by taking $k=2$ in Eq. \eqref{eq:RR},
\begin{equation}
\Psi_{\nsu{2}{2}} \left(\{z_{i}\}\right)= \mathcal{S}\left[ \prod_{i<j} \left(z_{i}^{1}-z_{j}^{1}\right)^2 \prod_{i<j} \left(z_{i}^{2}-z_{j}^{2}\right)^2 \right].
\label{eq:MooreRead}
\end{equation}
The wave function for each of the layers is a bosonic Laughlin state at $\nu=\frac{1}{2}$. 
Hence their ground state unit cells are given by $\left[01\right]$ and $\left[10\right]$. 
The ground state patterns for the full wave function, obtained by adding the occupation numbers for the two independent layers, are $\left[02\right],\left[20\right]$ and $\left[11\right]$. Inserting the fundamental quasielectron, the domain wall structures are given by \cite{Seidel2006abelian,Bergholtz2006pfaffian},
\begin{equation}
\begin{aligned}
\left[02\right]&\to\left[11\right]\\
 \left[20\right]&\to \left[11\right]\\
 \left[11\right]& \to \left[02\right]+\left[20\right].
\end{aligned}
\label{eq:DWall}
\end{equation}
Note that in the last case we can hop either electron in the [11] pattern, thus giving two fusion outcomes, which is the hallmark of nonabelian statistics. 
Further, we can again find the possible domain wall structure by the simple hopping rule described earlier, except that we now also allow doubly occupied orbitals.  

\begin{figure}
\includegraphics[width=48mm]{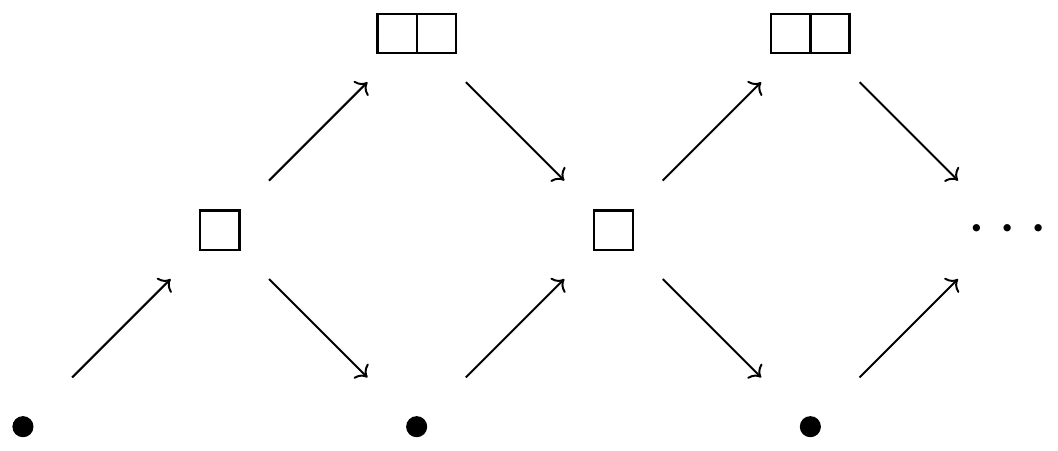}
\caption{Bratteli diagram for $\hat{\omega}_1$ in $\su{2}{2}$.}
\label{fig:Bratteli}
\end{figure}

The domain wall structures \eqref{eq:DWall} can be identified with the fusion rules of the fundamental representation of the algebra $\su{2}{2}$. 
The algebra $\su{2}{2}$ has three irreducible representations, which are written: $\left(2;0\right),\left(0;2\right)$, and $\left(1;1\right)$. We denote these in terms of Young tableaux as follows:
\begin{center}
$\underset{\left(2;0\right)}{{}^{\bullet}}\quad \quad \underset{\left(1;1\right)}{ \yng(1)} \quad \quad\underset{\left(0;2\right)}{ \yng(2)}$
\end{center}
Here the Young tableau of $\left(r_0;r_1\right)$ is the Young tableau associated with the $\su{2}{}$ representation $\lambda=r_1 \omega_1$, where $\omega_1$ is the fundamental representation of $\su{2}{}$. 
The fundamental representation of $\su{2}{2}$ is denoted $\hat{\omega}_1$ and it has the same Young tableau - a single box - as $\omega_1$ of $\su{2}{}$. 
We determine the fusion rules of $\hat{\omega}_1$ by virtue of the Littlewood-Richardson rule \cite{CFT}. 
These fusion rules are shown in  Fig. \ref{fig:Bratteli} as a Bratteli diagram, where each arrow indicates a fusion with $\hat{\omega}_1$. 
It is clear that this has exactly the same structure as the Bratteli diagram based on the domain wall structures Eq. \eqref{eq:DWall},  where each arrow corresponds to inserting a quasiparticle domain wall. 
For this, one identifies $\left[02\right]\leftrightarrow\left(2;0\right), \left[20\right]\leftrightarrow\left(0;2\right)$ and $\left[11\right]\leftrightarrow\left(1;1\right)$.  

Closely related to the algebra $\su{2}{2}$ is the Ising CFT $\pf{2}{2}{}$. In fact it has the same fusion rules as $\su{2}{2}$, although the conformal dimensions of the fields are different. The primary fields are commonly denoted $1,\psi$ and $\sigma$, and the fusion rules of $\sigma$ read
\begin{equation}
\begin{aligned}
1\times \sigma &= \sigma \\
\psi \times \sigma &= \sigma \\
\sigma \times \sigma &= 1+\psi .
\label{eq:Ising}
\end{aligned}
\end{equation} 
One can also identify $\left[02\right]\leftrightarrow 1$, $\left[20\right]\leftrightarrow\psi$, $\left[11\right]\leftrightarrow  \sigma$. We note, however, that for general $n$ and $k$, the domain wall structures will be identified with the fusion rules of generalizations of $\su{2}{2}$, rather than generalizations of the Ising CFT. 
\subsection{Nonabelian hierarchy states}
Let us now discuss  nonabelian hierarchy states of the form Eq. \eqref{eq:SU(n)_kSym}. 
For general $n$ and $k$ the wave function for a single layer has filling fraction $\nu=\frac{n}{n+1}$, so the ground state patterns in the thin torus limit are $\left[011\cdots1\right]=\left[01_n\right]$ and its translations \cite{TTLimit}. 
These are identical to the thin torus patters of the generalized Halperin states, except that the latter have an additional quantum number, namely the spin. 
In particular, the derivatives do not alter the thin torus patterns \footnote{On the (thin) cylinder, on the other hand, the corresponding patterns are different, but these differences are confined to the edges of the cylinder. }.
As the different layers in Eq.~\eqref{eq:SU(n)_kSym} are independent of each other, we obtain the ground state patterns of the fully symmetrized state by summing the occupation numbers of the patterns in each layer and discarding duplicates. 
Doing so one finds that the ground state degeneracy is ${n+k \choose k}$.
As before the quasiparticle domain wall structures are determined by a `hopping rule', i.e. the domain wall connects two sectors when the unit cell of the final sector can be obtained from the first by hopping an electron one site to the left. 
In the general case electrons can only be hopped into sites when the resulting occupancy is not greater than $k$. 
We will now show that the corresponding thin torus patterns are in one-to-one correspondence with the irreducible representations of $\su{n+1}{k}$. Furthermore, we show that the quasiparticle domain wall structures  correspond to the fusion rules of the fundamental representations $\hat{\omega}_1$ and $\hat{\omega}_n$ of $\su{n+1}{k}$. 

The irreducible representations of $\su{n+1}{k}$ read $\Lambda=\left(r_0;r_1,...,r_n\right)$, where the $r_\mu$ are non-negative integers, $\lambda=\sum_{i\geq1} r_i \omega_i$ is an $\su{n+1}{}$ representation in terms of the fundamental weights \footnote{We denote the fundamental weights of $\su{n_1}{}$ by $\omega_i$, and the fundamental weights (affine weights) of $\su{n+1}{k}$ by $\hat{\omega}_i$} $\omega_i$, and $r_0=k-\sum_{i\geq1} r_i$. There are ${n+k \choose k}$ irreducible representations. The map from the thin torus patterns onto these irreps is
\begin{equation}
\left[r_0 r_1 \cdots r_n\right] \mapsto \left(\bar{r}_0; \bar{r}_1,...,\bar{r}_n\right)
\label{eq:map}
\end{equation}
where $\bar{r}_\mu = k-r_\mu$ is particle-hole conjugation. 

The fundamental quasielectron corresponds to the pattern $\left[1\left(k-1\right)k\cdots k\right]$, which maps onto $\hat{\omega}_1=\left(\left(k-1\right);1,0,...,0\right)$ in $\su{n+1}{k}$ \footnote{Here, we have chosen the pattern $\left[0k_n\right]$ to correspond to the trivial representation}.
The corresponding Young tableau is a single box. Applying the Littlewood-Richardson rule, the fusion rules of $\hat{\omega}_1$ are described as follows.  Fusing $\hat{\omega}_1$ with an irrep $\left(r_0;r_1,...,r_n\right)$ yields a sum over irreps where one changes (imposing periodic boundary conditions) $r_\mu\to r_\mu-1$ and $r_{\mu+1}\to r_{\mu+1}-1$, as long as $0\leq r_\mu \leq k$. Translating this back into the thin torus patterns precisely corresponds to the `hopping rule' discussed earlier. Similarly, the fusion rules of $\hat{\omega}_n$ correspond to the `hopping rule' of the fundamental quasihole. 
This concludes the proof of the fusion rules for fractional quantum Hall states of type Eq. \eqref{eq:SU(n)_kSym}, but we can derive the fusion rules even for the more general states of type Eq. \eqref{eq:NAhier}. The proof for the latter hinges on the fact that the domain wall structure of an abelian state at filling fraction $p/q$ depends only on $q$. Thus, there is a one-to-one mapping of the domain wall patterns of $\Psi_{p/q}^k$ of  Eq. \eqref{eq:NAhier} to the domain wall patterns of $\Psi_{\left(q-1\right)/q}^k$ in Eq.~\eqref{eq:SU(n)_kSym} and the quasiparticle excitations of both states obey the same fusion rules. 

We note that we have thus far only considered domain wall structures involving the fundamental quasiparticles. As a result we find only a subset of the full set of fusion rules, namely those that involve the fusion of $\hat{\omega}_1$ and $\hat{\omega}_n$. However, the entire fusion algebra can be fixed by studying the domain wall structures of domain walls with a higher charge, which have the following `hopping rule': 
starting from an initial unit cell, the final unit cells are obtained by hopping $i$ electrons one site to the left, with the constraint that we can only hop 1 electron from each site. These rules map onto the fusion rules of the fundamental representations $\hat{\omega}_i$, which generate the entire fusion algebra \cite{GannonWalton}. Hence, the full fusion algebra is equal to $\su{n+1}{k}$. 

\section{Example; $n=3,k=2$}
\label{sec:ex}
To make the above discussion less abstract we consider an explicit example, taking $n=3$ and $k=2$ in Eq. \eqref{eq:SU(n)_kSym}. The simpler case $n=k=2$, related to the NASS wave function, was already studied in Ref. \cite{NAHierarchy}. The relevant nonabelian hierarchy wave function is given by 
\begin{widetext}
\begin{align}
\Psi_{3/4}^{2} \left(\left\{ z_i\right\}\right)&= \mathcal{S}\left[ \partial_2 \partial_{3}^{2}\prod_{a=1}^{2} \left(\prod_{\sigma=1}^{3} \prod_{i<j} \left(z_{i,\sigma}^{a}-z_{j,\sigma}^{a}\right)^2 \prod_{\sigma<\sigma'}^{3} \prod_{i,j} \left( z_{i,\sigma}^{a}-z_{j,\sigma'}^{a}\right)\right)\right]\nonumber\\
&=\mathcal{S}\left[ \Psi_{3/4} \Psi_{3/4} \right].
\label{eq:ExWf}
\end{align}
\end{widetext}
As seen in the previous section, the thin torus ground state patterns for the state with $n=3$ and $k=1$ are given by  $\left[0111\right],\, \left[1011\right], \, \left[1101\right]$ and $\left[1110\right]$. Therefore, the ground state sectors for the full wave function, symmetrizing over two layers, are 
\begin{equation*}
\begin{aligned}
&\left[0222\right],\left[2022\right],\left[2202\right],\left[2220\right],\\
&\left[1122\right],\left[2112\right],\left[2211\right],\left[1221\right],\\
&\left[1212\right],\left[2121\right].
\end{aligned}
\end{equation*}

\begin{figure}[h]
\includegraphics[width=80mm]{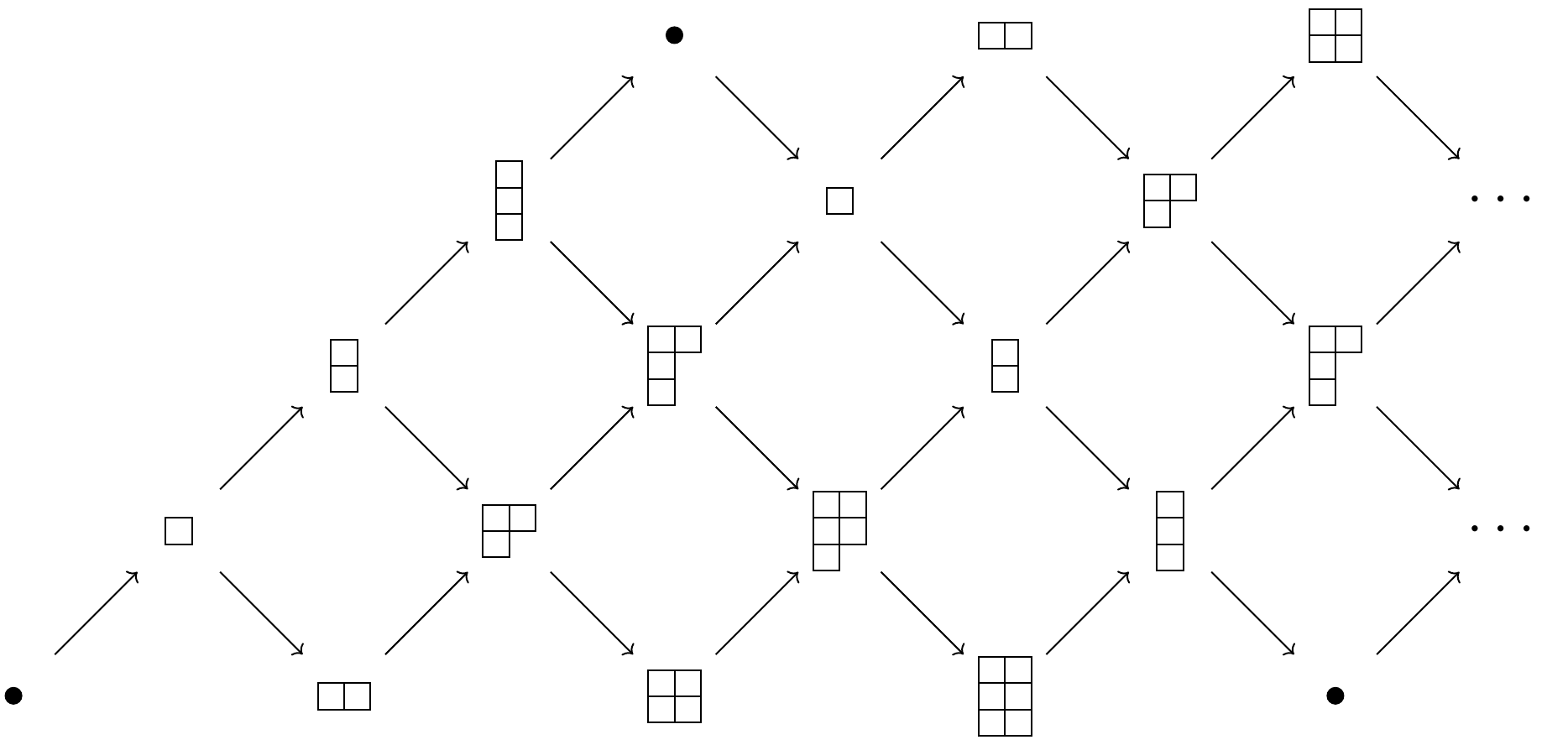}
\caption{Bratteli diagram for $\hat{\omega}_1$ in $\su{4}{2}$.}
\label{fig:Bratteli2}
\end{figure}

The domain wall structures involving the fundamental quasiparticle can be found by hopping electrons one site to the left, yielding
\begin{equation*}
\begin{aligned}
\left[0222\right] &\to \left[1122\right]\\
\left[1122\right] &\to \left[1212 \right]+\left[2022\right]\\
\left[1212\right]&\to \left[2112\right]+ \left[1221\right].\\
\end{aligned}
\end{equation*}
The remaining domain wall structures may be found by translating all `unit cells' simultaneously. Using the map  Eq. \eqref{eq:map}, this precisely corresponds to the fusion rules of $\hat{\omega}_1$ in $\su{4}{2}$, shown in Fig. \ref{fig:Bratteli2} as a Bratteli diagram. 

The clustered spin-singlet state corresponding to the given case has $n=3$ values for the pseudospin and is given by the conformal block
\begin{multline}
\Psi_{\nsu{4}{2}}\!\left(\left\{z_{i,\sigma}\right\}\right)= \\
\bigg<\!\prod_{i=1}^{M} V_{\tilde{\alpha}_1}\!\left(z_{i,1}\right) \prod_{i=1}^{M} V_{\tilde{\alpha}_2}\! \left(z_{i,2}\right)  \prod_{i=1}^{M} V_{\tilde{\alpha}_3} \!\left(z_{i,3}\right)\bg\!\bigg>,
\label{eq:ExSU4}
\end{multline}
where the electron operators are given by 
\begin{equation}
V_{\tilde{\alpha}} \! \left(z\right) = \psi_{\tilde{\alpha}} \!\left(z\right) {:}e^{ i \tilde{\alpha} \cdot \vec{\varphi} \left(z\right)/\sqrt{2}}{:},
\end{equation}
i.e. a product of parafermions $\psi_{\tilde{\alpha}}$ of $\pf{4}{2}{3}$ and vertex operators in terms of bosonic fields $\vec{\varphi}=\left(\varphi_1,\varphi_2,\varphi_3\right)$. The roots $\tilde{\alpha}_i$ of $\su{4}{}$ can be chosen as
\begin{equation}
\begin{aligned}
\tilde{\alpha}_1 &= \left(\frac{2}{\sqrt{3}},0,-\frac{2}{\sqrt{6}} \right)\\
\tilde{\alpha}_2 &=  \left(\frac{2}{\sqrt{3}},\frac{1}{\sqrt{2}},\frac{1}{\sqrt{6}}\right)\\
\tilde{\alpha}_3 &= \left( \frac{2}{\sqrt{3}},-\frac{1}{\sqrt{2}},\frac{1}{\sqrt{6}}\right).
\end{aligned}
\end{equation}
Here we have used the `charge representation' of the roots such that it is charge neutral in all but the first sector, i.e. the second components and the third components of the roots $\tilde{\alpha}_i$ sum to zero. The field $\varphi_1$ then carries the physical charge -- see Appendix \ref{sec:roots}.

Alternatively, the correlator Eq. \eqref{eq:ExSU4} can be expressed as
\begin{equation}
\Psi_{\nsu{4}{2}}\!\left( \left\{z_{i}\right\}\right) = \mathcal{S}_l \left[ \prod_{a=1}^{2} \Psi_{\nsu{4}{1}}^{a} \!\left(\left\{z_{i,\sigma}^{a} \right\}\right)\right],
\end{equation}
where $\mathcal{S}_l$ symmetrizes over the coordinates in the different pseudospin groups, i.e. it symmetrizes over the coordinates in pseudospin groups 1, 2, and 3 separately.

Returning to the nonabelian hierarchy wave function, it is related to the above wave function via Eq. \eqref{eq:SU(n)_kSym}, which yields
\begin{align}
\Psi_{3/4}^{2} \left(\left\{z_i\right\}\right)&= \mathcal{S} \left[ \partial_2 \partial_{3}^{2} \mathcal{S}_l \left[\prod_{a=1}^{2} \Psi_{\nsu{4}{1}}^{a}\! \left(\left\{z_{i}^{a}\right\}\right)\right] \right]\nonumber\\
& = \mathcal{S} \left< \prod_{i=1}^{M} \tilde{V}_{\tilde{\alpha}_1}\left(z_i\right) \prod_{i=1}^{M} \tilde{V}_{\tilde{\alpha}_2} \left(z_i\right) \prod_{i=1}^{M}\tilde{V}_{\tilde{\alpha}_3} \left(z_i\right) \right>,
\end{align}
where the tilde on the operators $V$ denotes that we have included the derivatives in the operator definition. 
The nonabelian hierarchy wave function is hence expressed as a symmetrizer over a correlator of electron operators related to $\su{4}{2}$.

The quasiholes are described by the operators 
\begin{align}
H_{\tilde{\omega}_i} \!\left(\eta\right) = \sigma_{\tilde{\omega}_i}\! \left(\eta\right) e^{i\tilde{\omega}_i\cdot \vec{\varphi} /\sqrt{2}},
\end{align}
where the fields $\sigma_{\tilde{\omega}_i} = \Phi^{\omega_3}_{\tilde{\omega}_i}$ are primary fields of the appropriate parafermion theory (see Appendix \ref{sec:para} for more details). These fields carry a label $\omega_3$ as well as a label $\tilde{\omega}_i$ which is given 
\begin{equation}
\begin{aligned}
\tilde{\omega}_1 &= \omega_1-\omega_2=\left(\frac{1}{2\sqrt{3}},0,-\frac{2}{\sqrt{6}} \right)\\
\tilde{\omega}_2 &= \omega_2-\omega_3 =  \left(\frac{1}{2\sqrt{3}},\frac{1}{\sqrt{2}},\frac{1}{\sqrt{6}}\right)\\
\tilde{\omega}_3 &=\omega_3 = \left( \frac{1}{2\sqrt{3}},-\frac{1}{\sqrt{2}},\frac{1}{\sqrt{6}} \right)
\end{aligned}
\end{equation} 
in the `charge representation'. The $\tilde{\omega}_i$ satisfy $\tilde{\alpha}_i \cdot \tilde{\omega}_j = \delta_{ij}$. 
These quasihole operators generate all quasihole states, both for the clustered spin-singlet states and for the nonabelian hierarchy states. 
In the former case, the operators are linearly independent, but in the latter case, where pseudospin is not any longer a good quantum number, they generically are not. 
The derivatives ensure that they are still distinguishable and, thus, have still the same statistics as their spin-singlet counterparts \cite{Hansson2017quantum}. 

\section{Conclusion and summary}
\label{sec:conclusion}

In this manuscript we have analyzed generalizations of the nonabelian hierarchy wave functions that were originally introduced in \cite{NAHierarchy}. 
We showed that the wave functions Eq. \eqref{eq:NAhier}, given by
\begin{align}
\Psi_{p/q}^k=\mathcal{S}\big[\underbrace{\Psi_{p/q}\ldots \Psi_{p/q}}_{k\, \text{times}} \big]
\end{align}
have a filling fraction $kp/q$, a topological ground state degeneracy of ${q+k \choose k}$, and harbor quasiparticle excitations with fractional charge $\pm e/q$ and $\su{q}{k}$ type fusion rules. 
For a subset of these states, namely those with $p=q-1$, we have determined the CFT description that should make all the topological properties, in particular the braiding properties of quasiparticles, manifest \cite{read2009nonabelian}. 

The relevant CFT is a product of Gepner parafermions and vertex operators of chiral bosons. Both are unitary and rational CFTs, which is a necessary condition for describing gapped, incompressible quantum liquids. However, this condition is not sufficient.
In fact, there are several examples of rational, unitary CFTs that give gapless wave functions, see e.g.  Refs. ~\cite{Simon2010quantum,Jackson2013entanglement}. 
In order to resolve this, one needs to study the wave functions numerically. 
 
The simplest examples of type Eq. \eqref{eq:NAhier}  have been already been studied numerically \cite{BipartiteCF,sreejith2013tripartite}, where it was found that they have a large overlap with the exact diagonalization ground state of certain $3$- and $4$-body clustering Hamiltonians.  
In order to determine the screening properties of the nonabelian wave functions (and thus whether or not they can describe gapped systems), one however needs to employ different numerical tools.

A promising method could be the one introduced by Zaletel and Mong \cite{Zaletel2012exact}, who showed that certain quantum Hall model wave functions can be written as exact matrix product states, utilizing the underlying CFT description. 
This method has been generalized to the Read-Rezayi series \cite{MPSlong}, where it was used to verify the braiding properties of the quasiholes of the $Z_3$ parafermion state numerically \cite{Wu2015matrix,Wu2014braiding} and to confirm the screening properties of a variety of different model states \cite{Estienne2015correlation}. It can also be generalized to describe quasielectron excitations \cite{MPSqe}. 
At least for the simpler examples of type Eq. \eqref{eq:NAhier}, the MPS description should be able to determine if the wave functions are in the screening phase and determine the braiding properties that should be manifest in our CFT description.

\section*{Acknowledgements}
We would like to thank Eddy Ardonne for many helpful discussions. M.H. acknowledges partial support through the Emmy-Noether program of the Deutsche Forschungsgemeinschaft under grant no. HE 7267/1-1. Y.T. and M.H. acknowledge support from the Deutsche Forschungsgemeinschaft under grant no. CRC183. 

\begin{appendix}
\section{Roots for the clustered spin-singlet states}
\label{sec:roots}
As mentioned in the text, the generalized Halperin states have an $\su{n+1}{1}$ symmetry. The low energy effective field theory for these states is given by
\begin{equation}
\mathcal{L}_{n} = \frac{1}{4\pi} \epsilon^{\mu \nu \rho} a_{\mu}^{i} \left(K_n\right)_{ij}\partial_\nu  a_{\rho}^{j}+\frac{1}{2\pi} t_i \epsilon^{\mu \nu \rho} A_{\mu} \partial_{\nu} a_{\rho}^{i},
\end{equation}
where the $a_{\mu}^{i}$ are Chern-Simons fields, $A_\mu$ is the external electromagnetic field, and $\left(K_n\right)_{ij} = 1+\delta_{ij}$ is the $K$-matrix \cite{WenZeeEffective}. The $K$-matrix is related to the Cartan matrix $A_n$ of $\su{n+1}{}$ by a similarity transformation. Explicitly, 
\begin{equation}
A_n=\left(\begin{array}{cccccc}
2 & -1 & 0 & 0 & \cdots & 0\\
-1 & 2 & -1 & 0 & \cdots & 0\\
0 & -1 & 2 & -1 & \cdots & 0\\
\vdots &  & \ddots & \ddots & \ddots & \vdots \\
0 & 0 & \cdots & -1 & 2 & -1\\
0 & 0 & \cdots & 0 & -1 & 2

\end{array}\right)
\label{eq:Cartmat}
\end{equation}
and $K_n=WA_nW^T$ for $W_{ij} = \delta_{ij} + \delta_{i+1\,j}$. Letting $\alpha_1,...,\alpha_n$ be simple roots of $\su{n+1}{}$ with inner products as encoded in the Cartan matrix Eq. \eqref{eq:Cartmat}, the set of roots
\begin{equation}
\tilde{\alpha}_i = \sum_{j=1}^{i} \alpha_j
\label{eq:roots}
\end{equation} 
have inner products as encoded in the $K$-matrix; $\tilde{\alpha}_i \cdot \tilde{\alpha}_j= K_{ij}$. Using these roots for the vertex operators Eq. \eqref{eq:AVertOp} one obtains the wave function $\Psi_{\nsu{n+1}{1}}$, Eq. \eqref{eq:ASUN}.

A simple explicit representation is obtained by choosing the vectors such that the $i$-th vector has only the first $i$ components non-zero and to fix the components by requiring the correct inner products. Hence $\tilde{\alpha}_1 = \left(\sqrt{2},0,...,0\right)$ and $\tilde{\alpha}_2 = \left(\sqrt{\frac{1}{2}},\sqrt{\frac{3}{2}},0,...,0\right)$, and so on. A more natural representation for the roots is the `charge representation', which follows from the above by means of a simple rotation. Letting $w=\sum_i \tilde{\alpha}_i$, one performs a rotation $O$ on $w$ such that it lies along the first axis, i.e. $Ow=\left(\sqrt{n\left(n+1\right)},0,...,0\right)$. Then, the roots $O\tilde{\alpha}_i$ obey charge neutrality in all but the first sector, so that the background charge operator $\bg$ need only depend on a single field. This field may then be associated with physical charge. Since $\left(O \tilde{\alpha}_{i}\right)^{1}=\sqrt{\frac{n+1}{kn}}$ for all $i$, each electron type carries the same charge equal to $\frac{1}{\sqrt{\nu}}$ with $\nu$ being the filling fraction of $\Psi_{\nsu{n+1}{k}}$. Note that the factor $\sqrt{\frac{1}{k}}$ comes from the definition of the vertex operators for $k>1$. 

The fundamental weights of $\su{n+1}{}$ are determined by the condition $\omega_i \cdot \alpha_j=\delta_{ij}$, where $\alpha_j$ are the simple roots. It follows that ${\omega}_{i}^1 = \frac{n+1-i}{n+1} q_e$ in the `charge representation', where $q_e=\sqrt{\frac{n+1}{kn}}$ is the electron charge. The relevant weights for the quasihole operators Eq. \eqref{eq:AQh} and Eq. \eqref{eq:NASSH} are the weights dual to the vectors $\tilde{\alpha}$. It is easy to show that $\tilde{\alpha}_i \cdot {\tilde{\omega}}_j=\delta_{ij}$ is solved by taking $\tilde{\omega}_i= \omega_i-\omega_{i+1}$ for $1\leq i\leq n-1$ and $\tilde{\omega}_n=\omega_n$. These all carry the same charge $\tilde{\omega}_i^1=\frac{1}{n+1} q_e$.

\section{Parafermions}
\label{sec:para}
We provide a brief introduction to parafermion CFTs, referring the reader to \cite{GepnerParafermions} for more details. We will focus on the parafermion theory based on $\su{n+1}{}$ for $n\geq 1$. The most relevant  primary fields of these conformal field theories are the parafermions, which appear in vertex representations of the WZW algebra $\su{n+1}{k}$ for $k>1$ \cite{WESS197195,WITTEN1983422}. It is these representations that we use construct the clustered spin-singlet states. The WZW current algebra reads
\begin{equation}
J^a\!\left(z\right)\! J^b\!\left(w\right) \sim \frac{k \delta_{ab}}{\left(z-w\right)^2} + \frac{f^{abc}J^c \!\left(w\right)}{\left(z-w\right)}.
\label{eq:WZW}
\end{equation}
At level $k=1$ we can represent this algebra by using \footnote{Here one also needs to introduce correction factors $c_\alpha$ for the vertex operators - see \cite{CFT} for more details} the vertex operators $V_{\tilde{\alpha}} \!\left(z\right)=e^{i\tilde{\alpha}\cdot \vec{\varphi}\left(z\right)}$ and the Cartan currents $H^i = i \alpha_i \cdot \partial \vec{\varphi} \left(z\right)$, where $\tilde{\alpha}$ and $\alpha_i$ are the roots and simple roots of $\su{n+1}{}$. 

For levels $k>1$ the vertex operators and Cartan currents need to be modified in order to yield the OPEs Eq. \eqref{eq:WZW}. An initial guess would be to take $H^i \!\left(z\right)= i \sqrt{k} \alpha_i \cdot \partial_z \vec{\varphi} \left(z\right)$ and $V_{\tilde{\alpha}} \!\left(z\right)= \sqrt{k}\exp \left[i\tilde{\alpha} \cdot \vec{\varphi} \left(z\right)/\sqrt{k} \right]$. However, although the Cartan currents yield the correct OPEs, the vertex operators have the wrong conformal dimension $\Delta_V=1/k$ instead of 1. This issue is solved by introducing a  parafermion $\psi_{\tilde{\alpha}}\!\left(z\right)$, so that
\begin{equation}
V_{\tilde{\alpha}} \!\left(z\right) = \sqrt{k} \psi_{\tilde{\alpha}}\! \left(z\right)  \exp \left[ i\tilde{ \alpha }\cdot\vec{ \varphi} \left(z\right) /\sqrt{k}\right].
\end{equation} 
Like the vertex operator, the field $\psi_{\tilde{\alpha}}$ is labeled by a root $\tilde{\alpha}$, and the OPEs of the parafermions read
\begin{equation}
\begin{aligned}
\psi_{\tilde{\alpha}}\! \left(z\right) \psi_{-\tilde{\alpha}}\! \left(w\right) &= \left(z-w\right)^{-2+2/k}\\
\psi_{\tilde{\alpha}}\! \left(z\right) \psi_{\tilde{\beta}}\! \left(w\right) & = K_{\tilde{\alpha},\tilde{\beta}}\left(z-w\right)^{1-\tilde{\alpha}\cdot \tilde{\beta}/k} \psi_{\tilde{\alpha}+\tilde{\beta}}\! \left(w\right),
\end{aligned}
\end{equation}
with $K_{\tilde{\alpha},\tilde{\beta}}$ some constants. This reproduces Eq. \eqref{eq:WZW} for $k>1$, and the vertex operator can be used to construct the clustered spin-singlet states -- we drop the proportionality factor $\sqrt{k}$ in that case.  

The full parafermion theory, written $\pf{n+1}{k}{n}$, contains other fields besides the parafermions. The primary fields in this theory are written $\Phi^{\Lambda}_{\lambda}$. Here $\Lambda=\left(\Lambda_1,...,\Lambda_n\right)$ consist of the last $n$ components of the affine weight $\hat{\Lambda} = \left(\Lambda_0;\Lambda_1,...,\Lambda_n\right)$. The label $\lambda$ is an element of the weight lattice, i.e. 
\begin{equation}
\lambda = \left(\lambda_1,\lambda_2,...,\lambda_n\right)=\sum_i \lambda_i \omega_i,
\end{equation}
where the $\lambda_i$ are integers. These labels are defined only modulo $k$ times the (long) root lattice $M_L$ of $\su{n+1}{k}$. That is $\Phi^{\Lambda}_{\lambda} = \Phi^{\Lambda}_{\lambda'}$ if $\lambda = \lambda'+k\alpha$, where $\alpha = \sum_i r_i \alpha_i$ is an element of the (long) root lattice spanned by the simple roots $\alpha_i$.

The primary fields are subject to the following conditions:
\begin{enumerate}
\item $\lambda$ must be obtainable from $\Lambda$ by adding or subtracting the simple roots.
\item In order for the theory to behave properly under modular transformations, we must identify $\Phi^{\left(\Lambda_1,...,\Lambda_n\right)}_{\left(\lambda_1,...,\lambda_n\right)} = \Phi^{\left(\Lambda_0,...,\Lambda_{n-1}\right)}_{\left(\lambda_1+k,\lambda_2,...,\lambda_n\right)}$. 
\end{enumerate}
The fusion rules between the primary fields are given by
\begin{equation}
\Phi^{\Lambda}_{\lambda} \times \Phi^{\Lambda'}_{\lambda'} = \sum_{\Lambda''} \Phi^{\Lambda''}_{\lambda+\lambda'  \mathrm{mod}\, kM_L}
\label{eq:fusion}
\end{equation}
where the sum extends over the $\Lambda''$ labels associated to the irreps of $\su{n+1}{k}$ that appear in the fusion rule $\hat{\Lambda} \times \hat{\Lambda}'$. Moreover, one may have to impose field identifications on the primary fields that appear.

A simple example is the Ising CFT $\pf{2}{2}{}$. The irreps of $\su{2}{2}$ are $\left(2;0\right),\left(1;1\right),\left(0;2\right)$, so the $\Lambda$ labels are $\Lambda=0,1,2$. Since $\su{2}{}$ has only one simple root $\alpha=2$, condition 1. implies that only those $\lambda$ with the same parity as $\Lambda$ are allowed. Since the $\lambda$ labels are only defined modulo $k=2$ times the root lattice, i.e. modulo 4, one obtains a set of six fields; $\Phi^{0}_{0},\Phi^{0}_{2},\Phi^{1}_{1},\Phi^{1}_{3},\Phi^{2}_{0}$, and $\Phi^{2}_{2}$. Condition 2. leaves only the fields $\Phi^0_0=1$, $\psi=\Phi^0_2$ and $\sigma=\Phi^1_1$. The fusion rules yields the Ising fusion rules, some of which are given in Eq. \eqref{eq:Ising}.

\paragraph*{The parafermion theory $\pf{4}{2}{3}$}
We discuss briefly the parafermion theory $\pf{4}{2}{3}$, relevant for the example in Section \ref{sec:ex}. In this case, the $\Lambda$ labels are 
\begin{equation}
\begin{aligned}
\left(0,0,0\right),\left(2,0,0\right),\left(0,2,0\right),\left(0,0,2\right)&\\
\left(1,0,0\right),\left(1,1,0\right),\left(0,1,1\right),\left(0,0,1\right)&\\
\left(0,1,0\right),\left(1,0,1\right)&.
\end{aligned}
\end{equation}
The allowed labels $\lambda$ can be determined by adding to the $\Lambda$ labels the simple roots $\alpha_1=\left(2,-1,0\right),\alpha_2=\left(-1,2,-1\right)$ and $\alpha_3=\left(0,-1,2\right)$ and their sums. Invoking the field identifications, one is left with $20$ primary fields. Of these, the most relevant are the parafermions
\begin{equation}
\begin{aligned}
\psi_{\alpha_1} &= \Phi^{\left(0,0,0\right)}_{\alpha_1}\\
\psi_{\alpha_1+\alpha_2}&=\Phi^{\left(0,0,0\right)}_{\alpha_1+\alpha_2}\\
\psi_{\alpha_1+\alpha_2+\alpha_3}&=\Phi^{\left(0,0,0\right)}_{\alpha_1+\alpha_2+\alpha_3}
\end{aligned}
\end{equation}
used to construct the electron operators. The spin fields that are used to construct the quasihole operators Eq. \eqref{eq:NASSH} are 
\begin{equation}
\begin{aligned}
\sigma_{\tilde{\omega}_1}&= \Phi^{\omega_3}_{\omega_1-\omega_2}\\
\sigma_{\tilde{\omega}_2}&=\Phi^{\omega_3}_{\omega_2-\omega_3}\\
\sigma_{\tilde{\omega}_3}&=\Phi^{\omega_3}_{\omega_3}.
\end{aligned}
\end{equation}
Here $\omega_3=\left(0,0,1\right)$ corresponds to the irrep  $\hat{\omega}_3 = \left(1;0,0,1\right)$ of $\su{4}{2}$.

\section{Quasiparticle degeneracy}
\label{sec:qpdeg}
The fusion rules for the nonabelian hierarchy states $\Psi_{n/n+1}^{k}$ are those of $\su{n+1}{k}$, which are nonabelian for $k>1$. A necessary property of nonabelian excitations is a nontrivial degeneracy when $N$ such particles are inserted at well-separated positions. For the Moore-Read state Eq. \eqref{eq:MooreRead} this degeneracy is given by $2^{N/2 - 1}$ for $N>0$, which is encoded in the Bratteli diagram for $\su{2}{2}$, Fig. \ref{fig:Bratteli}. Namely, the $N$ quasiparticles must fuse to the identity in order to yield a non-zero wave function: the degeneracy is given by the number of distinct paths of length $N>0$ in the Bratteli diagram from the identity to the identity. This is easily seen to be $2^{N/2-1}$, when $N$ is a multiple of two.

The quasiparticle degeneracy for the Read-Rezayi states, i.e. for the algebras $\su{2}{k}$, was computed in \cite{Ardonne2008degeneracy}.  For $\su{2}{k},$ the representations $\left(r_0;r_1\right)$ appearing in the vertices in the Bratteli diagram may be labeled by $r_1=0,...,k$. The number of paths going from $r_1$ to $r_1'$ when fusing with $\hat{\omega}_1$ $N$ times, which we will denote by $d_{2,k} \left(r_1,r_1',N\right)$, is given by  $d_{2,k}\left(r_1,r_1',N\right)=\left(N_{1}\right)_{r_1,r_1'}^{N}$. That is, the degeneracy is the $r_1,r_1'$ matrix element of the $N$th power of the fusion matrix $N_{1}$ of $\hat{\omega}_{1}$ in $\su{2}{k}$. By virtue of the Verlinde formula, this may be written in terms of the matrix elements of the modular $S$ matrix of $\su{2}{k}$. The result is
\begin{widetext}
\begin{equation}
d_{2,k}\left(r_1,r_1',N\right)=\frac{2}{k+2}\sum_{m=0}^{k}\sin\left(\frac{\left(r_1+1\right)\left(m+1\right)\pi}{k+2}\right)\sin\left(\frac{\left(r_1'+1\right)\left(m+1\right)\pi}{k+2}\right)\left(2\cos\left(\frac{\left(m+1\right)\pi}{k+2}\right)\right)^N.
\end{equation}
\end{widetext}
Here we have used that
\begin{eqnarray}
\left(N_{1}\right)_{r_1,r_1'}^{N}&=&\sum_{m=0}^{k}S_{r_1\,m}S_{r'_1\,m}\left(\frac{S_{1m}}{S_{0m}}\right)^{N}\\
S_{r_1,r_1'}&=&\sqrt{\frac{2}{k+2}}\sin\left(\frac{\left(r_1+1\right)\left(r_1'+1\right)\pi}{k+2}\right)
\end{eqnarray}

We now turn to the nonabelian hierarchy states. By virtue of the rank-level duality \cite{CFT}, the quasiparticle degeneracy of the nonabelian hierarchy state based on $\su{n+1}{2}$, i.e. at level $k=2$, is closely related to the quasiparticle degeneracy of $\su{2}{n+1}$. In particular, the Bratteli diagrams of $\hat{\omega}_{1}$ and $\hat{\omega}_{n}$ in $\su{n+1}{2}$ have exactly the same structure as the Bratteli diagrams of $\hat{\omega}_{1}$ in $\su{2}{n+1}$, although the labeling of the vertices is different. As a result, we may deduce the number of paths from identity to identity in the Bratteli diagram for $\su{n+1}{2}$ by counting the paths between representations in the same positions in the Bratteli diagram for
$\su{2}{n+1}$. In the Bratteli diagram for $\su{n+1}{2}$, the identity appears at the positions $\left(s\left(n+1\right),n+1\right)$ for $s$ an odd integer and $\left(s\left(n+1\right),0\right)$ for $s$ even. On those positions in the Bratteli diagram for $\su{2}{n+1}$, we find the representation $\left(0;n+1\right)$ for $s$ odd and the identity for $s$ even.
As a result, $d_{n+1,2} \left(0,0,N=\left(n+1\right)s\right)$ is equal to $ d_{2,n+1}\left(0,\left(n+1\right)\left(s\, \mathrm{mod} 2\right),N=\left(n+1\right)s\right)$, so that
\begin{widetext}
\begin{equation}
d_{n+1,2}\left(0,0,N=\left(n+1\right)s\right)=\frac{2}{n+3}\sum_{m=0}^{n+1} \left(-1\right)^{m\left(s\,\mathrm{mod} 2\right)}\sin ^2\left(\frac{\left(m+1\right)\pi}{n+3}\right)\left(2\cos\left(\frac{\left(m+1\right)\pi}{n+3}\right)\right)^N,
\end{equation}
\end{widetext}
and $d_{n+1,2} \left(0,0,N\right)$ is zero if $N\neq \left(n+1\right)s$. 
For the general case $\su{n+1}{k},$ the quasiparticle degeneracy can be found
by using the Verlinde formula with the general expression for the modular $S$ matrix for $\su{n+1}{k}$ \cite{CFT}, although the expressions become too complicated to provide simple counting formulas in most cases.  
Two simple examples are  $\su{3}{3}$ and $\su{4}{4}$ where the degeneracy is given by
\begin{equation}
\begin{aligned}
d_{3,3}\left(0,0,N=3s\right)&=\frac{1}{12}\left(3 \delta_{s,0}+8^{s}+8\cos\left(\frac{\pi s}{3}\right)\right)\\
d_{4,4}\left(0,0,N=4s\right)&=\frac{1}{32}\bigg\{6\delta_{s,0}+2^{2s+1}\left(1+\left(-1\right)^{s}\right)\\
&+8^{s}\left[\left(3+2\sqrt{2}\right)^{s-1}+\left(3-2\sqrt{2}\right)^{s-1}\right]\\
+8\cos\left(\frac{\pi s}{4}\right)&\left.\left[\left(2+\sqrt{2}\right)^{2s-1}+\left(-1\right)^{s}\left(2-\sqrt{2}\right)^{2s-1}\right]\right\}
\end{aligned}
\end{equation}

\end{appendix}
\bibliography{ON}
\end{document}